\documentstyle[preprint,prb,aps,epsf]{revtex}
\tightenlines
\def\rect#1#2{{\vcenter{\vbox{\hrule height.3pt
            \hbox{\vrule width.3pt height#2truecm \kern#1truecm
            \vrule width.3pt}
            \hrule height.3pt}}}}
\def\square{\rect{0.15}{0.15}}

\def\inseps#1#2{\def\epsfsize##1##2{#2##1} \centerline{\epsfbox{#1}}}
\begin{document}
\draft
\title{Monte Carlo study of the growth of striped domains}
\author{E.N.M. Cirillo$^1$, G. Gonnella$^1$, 
S. Stramaglia$^2$}
\address{$^1$Dipartimento di Fisica dell'Universit\`a di Bari and
Istituto Nazionale di Fisica Nucleare, Sezione di Bari, via Amendola
173, I-70126 Bari, Italy\\
$^2$Istituto Elaborazione Segnali ed
Immagini C.N.R., via Amendola 166/5, I-70126 Bari, Italy}
\date{\today}
\maketitle
\begin{abstract}
We analyze the dynamical scaling behavior
in a two-dimensional spin model with competing interactions after a quench
to a striped phase. We measure the 
growth exponents studying 
the scaling of the interfaces and the scaling of the shrinking time of a ball
of one phase plunged into the sea of another phase. Our results confirm
the predictions found in previous papers.
The correlation functions measured in the direction parallel and transversal
to the stripes are different as suggested by the existence of different
interface energies between the ground states of the model. 
Our simulations show anisotropic features for the correlations 
both in the case of single-spin-flip and spin-exchange dynamics. 
\end{abstract}


\section{Introduction}
When a system in a disordered phase is rapidly quenched to a temperature below
its ordering temperature it orders kinetically [\onlinecite{BR,G}]. 
In the early stage the 
fluctuations in the initial state are amplified and domains of macroscopic size 
are formed. At late time the ordering system is usually characterized by a 
single time-dependent length, the average domain size $R(t)$, which grows as a 
power law $R\sim t^{\gamma}$ (scaling regime) [\onlinecite{B}]. 
The growth exponent depends on the 
mechanism driving the phase separation. For a system with a scalar 
not-conserved order parameter (model A) the growth is curvature driven and 
$\gamma={1\over 2}$ [\onlinecite{LU}]. 
The mechanism in systems with conserved order 
parameter (model B) is the diffusion of the order 
parameter from interfaces of high curvature to 
regions of low curvature and $\gamma={1\over 3}$ [\onlinecite{LD,Leb}].
\par
This work concerns the ordering in the Ising version of the bidimensional
isotropic eight vertex model after a quench to the striped 
phase.
The reduced hamiltonian of the model is 
\begin{equation}
-\beta H = J_1 \sum_{<ij>} s_i s_j + J_2\sum_{<<ij>>} s_i s_j
          +J_3 \sum _{[i,j,k,l]} s_i s_j s_k s_l,
\label{eq:ham}
\end{equation}
where $s_i$ are Ising spins on a bidimensional square lattice and the sums 
are respectively on nearest neighbor pairs of spins, next to the nearest 
neighbor pairs and plaquettes. We have considered the case of periodic
boundary conditions.
\par
The quenching of model (\ref{eq:ham})
to the ferromagnetic phase with $J_3=0$ has been 
studied in both two and three dimensions [\onlinecite{SH}].  
When $J_2 <0$ there are energy barriers for the coarsening 
of domains, hence the 
system does not relax to equilibrium if the temperature is zero. These energy 
barriers do not increase with the linear dimension of domains in $D=2$, while 
in $D=3$ the energy barrier is proportional to the linear dimension of domains 
thus leading to a very slow dynamics [\onlinecite{SH}].
\par
The phase diagram of model (\ref{eq:ham}) when 
$J_2 < 0$, $|J_1| < 2|J_2|$ and for $J_3$ small 
is characterized by a critical curve 
separating the paramagnetic phase and the superantiferromagnetic 
phases (see e. g. [\onlinecite{BAX}]). In the 
SAF case the model has four ground states consisting of alternate plus and 
minus rows (columns). The order parameter for SAF phases is the difference 
between the magnetization on odd and even rows (columns).
\par
Quenching in the SAF phase, for model (\ref{eq:ham}) in $D=2$, has been
studied in [\onlinecite{SB}], where it has been shown that also in the striped 
phase the average size of domains grows as $R\sim t^{\gamma}$ where
$\gamma=\frac{1}{2}$ or $\gamma=\frac{1}{3}$ for single-spin-flip or
spin-exchange dynamics. 
\par
In a previous paper [\onlinecite{CGS}] we have shown
that the correlation functions measured in the direction parallel and
transversal to the stripes are different as suggested by the existence
of different interface energies between the ground states of the model.
Furthermore we furnished an explicit example of how the scaled correlation
functions can depend on the details of the system (see also [\onlinecite{RU}]).
In this paper we complete the analysis described in [\onlinecite{CGS}]. 
\par
In the next Section we concentrate on the scaling exponents and we
confirm the values found in [\onlinecite{SB}] by studying the scaling of the
amount of interfaces and the scaling of the shrinking time of a ``ball"
of one phase plunged into a sea of a different phase. 
\par
In Section 3 we study the asymptotic correlations in the model. As to the
single-spin-flip dynamics we show that the analysis of the pair correlation
functions introduced in [\onlinecite{SB}] gives results indistinguishable from
those obtained considering the longitudinal and transversal correlation
functions as described in [\onlinecite{CGS}]. As to the spin-exchange dynamics
we find an anisotropic behavior of the scaled correlation functions
as well. 
\par
Section 4 is devoted to the conclusions.
\vskip 2 cm

\section{Different ways to measure growth exponents}
The growth exponents of domains after a quenching in the SAF phase
have been already measured in [\onlinecite{SB}]; it is interesting to
measure those exponents studying the scaling properties of physical
quantities different from those considered in [\onlinecite{SB}]. 
\par
In order to describe the main features of domains we remark that
the ground state of the hamiltonian (\ref{eq:ham}) 
is four-fold degenerate, so the typical configuration of 
the system in the scaling regime consists of a patchwork of domains of four 
types (see Fig. 1a). In the following we refer 
to domains having magnetized rows 
as ``horizontal" domains, we call ``vertical" domains those with magnetized 
columns.
\par 
First of all we studied the amount of 
interfaces (see Fig. 1b) present in the system to monitor the  
domain growth towards the equilibrium, that is to estimate the dimension
of the domains. Indeed in two dimensions the total length of interfaces
per unit volume scales as the inverse of the average size of domains
(see [\onlinecite{HB}] where the scaling of interfaces is used to measure
the growth exponent in the case of ferromagnetic Ising model). 
\par
The definition of interfaces between different domains
is not straightforward in the case of the superantiferromagnetic phase.
In order to detect if a given site belongs to a domain or to the interface,
we compare the configuration of the system in a neighborhood of the site
with two given patterns in the following way: 
let us consider a site of the lattice, we denote it by the pair $(i,j)$, 
where $i$ 
and $j$ are respectively the row and column index, 
and we denote by $s(i,j)$ the corresponding spin. We consider
a positive integer number $L$, denote by $B(i,j)$
the $(2L+1)\times (2L+1)$ square block centered at site $(i,j)$ and 
compute the two following 
quantities 
\begin{equation}
\begin{array}{lll}
d_h &=&\sum_{k,l=-L}^L \left(1-\delta\left(s(i+k,j+l),{(-1)}^k s(i,j)\right) 
\right)\\
&&\\
d_v &=&\sum_{k,l=-L}^L \left(1-\delta\left(s(i+k,j+l),{(-1)}^l s(i,j)\right) 
\right)\\
\end{array}
\label{eq:dist}
\end{equation}
where $\delta (\cdot\; ,\; \cdot)$ is Kroneker's delta. We remark that
$d_h$ and $d_v$ provides, respectively, the ``distance" 
in $B(i,j)$ 
between the present configuration and a SAF horizontal or vertical domain.
If $d_h$ or $d_v$ are less than a fixed integer number $M$, we say that
the site $(i,j)$ belongs, respectively, to an horizontal or vertical
SAF domain; otherwise we say that the site $(i,j)$ is an interface site.
All the results that will be described in this paper have been obtained
with $L=1$ and $M=2$; we have checked that our results do not depend
on the choice of the parameters $L$ and $M$. 
\par
In Fig. 2 
it is depicted the logarithm of the total number $A$ of interface 
sites for a $512\times 512$ lattice system as a function of the logarithm 
of the time in the case of finite temperature
($\beta=1$) and parameters $J_1=0.1$, $J_2=-1$ and $J_3=0.1$.
Black circles and black squares (above and below in the picture) are Monte
Carlo results obtained averaging over $50$ different histories respectively
in the case of single-spin-flip [\onlinecite{GLA}] and spin-exchange dynamics
[\onlinecite{KAW}].
We find the scaling law $A\sim 
t^{-\gamma}$ with $\gamma=\frac{1}{2}$ and $\gamma=\frac{1}{3}$ respectively
in the case of single-spin-flip and spin-exchange dynamics, thus confirming 
the growth exponent found in [\onlinecite{SB}].
\par
The scaling law and the values of the exponent $\gamma$ have been shown
not to depend on the J's and on the value of $\beta$. In the case of
single-spin-flip dynamics the case of zero temperature [\onlinecite{zero}]
($\beta=\infty$) has been considered, as well.
We also checked that these results are not affected by finite size effects.
\par
Another way of measuring the exponent $\gamma$, in the case of not-conserved 
dynamics, consists in studying
the shrinking of a $L\times L$ square domain (ball) of one phase 
plunged into the sea of one of the three other phases. In Fig. 3 
it is shown the shrinking of a $41\times 41$ square vertical domain plunged
into the sea of the horizontal superantiferromagnetic phase. The contraction
of the ball is achieved via the corner erosion, indeed in this way no 
energy barrier must be bypassed. In [\onlinecite{BR}] it
has been remarked that the shrinking time $\tau$ of such a domain scales
as $\tau\sim L^{1\over\gamma}$. 
We found that the time of 
shrinking scales as $\tau \sim L^2$, see Fig. 4, 
for large $L$ and for any 
choice of the phase of the background, this confirms  
the growth exponent $\gamma={1\over 2}$.  
\vskip 2 cm 

\section{Anisotropic behavior of correlation functions}
Let us now turn to consider the correlation properties of growing domains in 
the scaling regime. In [\onlinecite{CGS}] we introduced two different types 
of correlation functions, 
the {\it longitudinal} and the {\it transverse} one; these two functions are
denoted respectively by $C_{\ell}(r)$ and $C_t(r)$ and they are defined as 
follows  
\begin{equation}
\begin{array}{lll}
C_{\ell} (r,t) &=&\big< s\left( i,j\right) s\left( i+\epsilon (i,j) r\; ,\; 
j+(1-\epsilon(i,j)
) r\right) \big>\\
&&\\
C_t (r,t) &=&\big< {(-1)}^r s\left( i,j\right) s\left( i+(1-\epsilon (i,j)) r
\; ,\;
j+ \epsilon(i,j)
    r\right) \big>\\
\end{array}
\label{eq:cl}
\end{equation}
where $\epsilon (i,j)$ is one (zero) if site $(i,j)$ belongs to a horizontal 
(vertical) domain and the average is performed over sites belonging 
to domains at time t (interface 
sites are excluded) and over different stories of the system. We remark that
the longitudinal correlation function measures the correlation properties of 
domains in the direction where spins are aligned, while the transverse 
correlation function measures the correlation properties in the direction where 
the spins are alternate.
\par
We found that the correlation functions introduced above have a scaling 
behavior in the scaling regime; in the case of single-spin-flip dynamics 
they depend only on the combination $r/\sqrt{t}$:
\begin{equation}
C_{\ell} (r,t) = f_{\ell} \left( {r\over \sqrt{t}}\right) \;\;\;\;\;\;\;\;
C_t (r,t) = f_t \left( 
{r\over \sqrt{t}}\right)
\label{eq:sca}
\end{equation}
Simulations discussed in [\onlinecite{CGS}] have shown that the scaling 
functions $f_{\ell}$ and $f_t$ are different,  
due to the presence of interfaces between the four types of domains 
having different interface energies.
The OJK theory [\onlinecite{OJK}] well describes the pair scaling functions
in anisotropic cases if a free parameter is used to fit the data in the
different directions (see [\onlinecite{CGS}]). 
\par
We also note that in this case two typical lengths characterize the 
ordering system, the average longitudinal domain size $R_{\ell}$ and 
the transverse
one $R_t$;
we evaluated them as 
\begin{equation}
R_{\ell}^{-1}={\sum k S_{\ell} (k)\over {\sum S_{\ell}(k)}} \;\;\;\;\;\;\;\;
R_t^{-1}={\sum k S_t (k)\over {\sum S_t(k)}}
\label{eq:fouri}
\end{equation}
where $S_{\ell} (k)$ and $S_t (k)$ are the Fourier 
transforms of $C_{\ell} (r)$ and 
$C_t (r)$ 
respectively. We verified that $R_{\ell}$ and $R_t$ are 
different but they both have
the correct scaling $R_{\ell} \sim t^{1\over 2}$ and $R_t \sim t^{1\over 2}$. 
\par
Now we check that 
the anisotropic behavior of domains growth does not depend on the peculiar
way in which (\ref{eq:cl}) measure the correlations, indeed we show that 
the analysis of the pair correlation functions
introduced in [\onlinecite{SB}] yields similar results. 
\par
As in [\onlinecite{SB}] let us divide the lattice into four sublattices
corresponding to a $2\times 2$ unit cell; a two-components order parameter
$\Psi=\{\psi_a\}_{a=1,2}$ can be expressed in terms of the sublattices
magnetizations $m_1$, $m_2$, $m_3$ and $m_4$, where
$m_{\nu}=\frac{1}{N}\sum_{i\in\nu}s_i$ and $N$ is the total number of 
sites in the lattice, by defining
\begin{equation}
\psi_1=m_1+m_2-m_3-m_4\;\;\;\;\;\;\;\;\;\psi_2=m_1-m_2-m_3+m_4
\;\;\; .
\label{eq:bin}
\end{equation}
The four ground states of model (\ref{eq:ham}) correspond to the four
following values of the order parameter $\Psi$:
\begin{equation}
{1\choose 0}\;\;\;{-1\choose 0}\;\;\;{0\choose 1}\;\;\;{0\choose -1}
\;\;\; .
\end{equation}
We observe that $|\psi_1 |=1$ ($|\psi_2 |=1$) corresponds to a striped
phase in the direction parallel to the $x$ ($y$) axis.
\par
By decomposing $\Psi$ into its contributions from individual unit cells
$\Psi_{\alpha}$, where $\alpha$ is the cell index, one can consider
the correlation between the components of the two-valued order parameter.
Two of such correlations can be put in correspondence with the longitudinal
and transverse correlation functions (\ref{eq:cl}). Indeed, let us consider
two cells $\alpha$ and $\alpha'$ connected by a vector parallel to the
$x$ axis: it is easy to see that the average of the product
$\psi_{\alpha,1}\psi_{\alpha',1}$ measures the longitudinal
correlations, while the averaged $\psi_{\alpha,2}\psi_{\alpha',2}$ is
a measure of the transverse correlations in the growing system. Conversely,
given two cells connected by a vector parallel to the
$y$ axis, $\psi_{\alpha,1}\psi_{\alpha',1}$ and 
$\psi_{\alpha,2}\psi_{\alpha',2}$ measure respectively the transverse
and the longitudinal correlations. In this framework the natural way
to define transverse and longitudinal correlation functions is the following:
\begin{equation}
\begin{array}{cc}
{\Gamma}_{\ell}(r,t)&=\frac{1}{2}\big<\psi_{\alpha,1}
                            \psi_{\alpha+r {\vec x},1}\big>
                     +\frac{1}{2}\big<\psi_{\alpha,2}
                            \psi_{\alpha+r {\vec y},2}\big>\\
{\Gamma}_{t}(r,t)   &=\frac{1}{2}\big<\psi_{\alpha,1}
                            \psi_{\alpha+r {\vec y},1}\big>
                     +\frac{1}{2}\big<\psi_{\alpha,2}
                            \psi_{\alpha+r {\vec x},2}\big>
\end{array}
\label{eq:cl-bin}
\end{equation}
where $\vec x$ and $\vec y$ are the unit vectors in the $x$ and $y$ directions
and the average is performed over all the cells at time t and over
different stories of the system.
\par
Although $C$'s and ${\Gamma}$'s are defined in a slightly different way, we 
note that from the physical point of view they are expected to satisfy
the following identities
\begin{equation}
{\Gamma}_{\ell}(r,t)=C_{\ell}(2r,t)\;\;\;\;\;
{\Gamma}_{t}(r,t)=C_{t}(2r,t)\;\;\; ,
\label{eq:ident}
\end{equation}
where the factor $2$ in the argument of the $C$'s is due to the fact that
the dynamical variables $\Psi_{\alpha}$ live on a lattice with spacing
which is twice the original one.
\par
In Fig. 5 we show the scaling collapse of the 
correlation functions
${\Gamma}_{\ell}$ and ${\Gamma}_{t}$. Data in Fig. 5 
above (below) have been obtained
by Monte Carlo simulations performed on a $400\times 400$ lattice system with
a single-spin-flip dynamics, at zero temperature and parameters $J_1=0.1$
($J_1=-0.1$), $J_2=-1$ and $J_3=0$. The correlation functions 
(\ref{eq:cl-bin}), taken at 
different times, have been plotted in terms of the scaling variable 
$z=r/\sqrt{t}$. From the picture it is clear that the anisotropy is
confirmed and that a simmetric behavior with respect to the change of the
sign of $J_1$ is observed.
\par
Solid lines in Fig. 5 represent the best fit of our data 
obtained with the OJK function
\begin{equation}
f(z)=\frac{2}{\pi}\sin^{-1}[\exp (-z^2/D)]
\;\;\; ;
\label{eq:ojk}
\end{equation}  
the optimal choices of the parameter $D$ is
$D_{\ell}=1.35$ $(1.15)$ and $D_t=1.15$ ($1.35$) respectively for
the longitudinal and transverse function in the case $J_1=0.1$ ($-0.1$).
The validity of the identities (\ref{eq:ident}) is confirmed by the fact
that the best fits of the $D$'s, in the case of the $C$'s (see 
[\onlinecite{CGS}]), are 
close to be four times the best fits of the $D$'s in the case of the
${\Gamma}$'s functions.
\par 
Finally we discuss the simulations we have performed with 
spin-exchange dynamics. 
We used heat bath spin-exchange between nearest neighbor 
pairs of spins [\onlinecite{KAW}]. 
In Fig. 6 the collapse of correlation functions (\ref{eq:cl}) is
shown: Monte Carlo data have been plotted in terms of the scaling
variable $z=r/t^{\frac{1}{3}}$. Simulations have been performed
on a $800\times 800$ lattice, at inverse temperature
$\beta=1$ and parameters $J_1=0.4$, $J_2=-1$ and $J_3=0$. We see that
the anisotropic behavior is observed in the case of the spin-exchange
dynamics as well. Again we found that the longitudinal and the transverse
correlation functions are exchanged when $J_1\rightarrow -J_1$. 

\section{Conclusions}
In this paper we have further analyzed the dynamical scaling behavior
in a two-dimensional spin model with competing interactions which
had been already investigated in [\onlinecite{SB,CGS}]. We measured the 
growth exponents studying physical magnitudes not considered in
previous papers: 
the scaling of the interfaces and the scaling of the shrinking time of a ball
of one phase plunged in the sea of another phase. Our results confirm
the predictions found in [\onlinecite{SB,CGS}].
\par
The anisotropic behavior of the correlations has been further investigated
by measuring the pair correlation functions introduced
in [\onlinecite{SB}] and their physical 
equivalence with  the correlation
functions used in [\onlinecite{CGS}] has been pointed out.
Our simulations show anisotropic features for the correlations in growing
system and the simmetric behavior with respect to the change of the sign
of the nearest neighbors coupling.
Finally, we have shown that the
anisotropy is observed also in the case of spin exchange dynamics.

\newpage
\par\noindent
{\bf Figure Captions}
\vskip 1 cm
\par\noindent
Fig 1: (a) Typical configuration of model (\ref{eq:ham}) in the scaling regime. 
Black (white) squares represent plus (minus) spins. The picture has been
obtained in a $100\times 100$ square lattice, at zero temperature, after
$150$ MCS (Monte Carlo Steps per site). 
(b) The same configuration as in Fig. 1a is depicted. Black squares 
represent the interface sites and grey (white) squares represent plus (minus)
spins belonging to domains (see the definitions in the text). 
\vskip 0.5 cm
\par\noindent
Fig 2: The logarithm of the number $A$ of interface sites is plotted versus
the logarithm of the time $t$ (number of iterations) in a $512\times 512$
square lattice. 
Black circles (above) and black squares (below) 
are Monte Carlo results
obtained averaging over $50$ different histories respectively in the case
of single-spin-flip and spin-exchange dynamics.
In both cases the inverse temperature is $\beta=1$ and the parameters
of the model have been chosen as follows: $J_1=0.1$, $J_2=-1$ and
$J_3=0.1$. The slope of the solid line is $-\frac{1}{2}$ above
and $-\frac{1}{3}$ below.
\vskip 0.5 cm
\par\noindent
Fig. 3: The shrinking of a $41\times 41$ vertical domain plunged into
the horizontal SAF phase is shown. Our simulation has been performed on
a $100\times 100$ square lattice, at zero temperature and with parameters
$J_1=0.1$, $J_2=-1$ and $J_3=0$. Black (white) squares represent plus (minus)
spins; from the left to the right and from the top to the bottom
configurations are shown at times $t=0,50,100,150,200,225$.
\vskip 0.5 cm
\par\noindent
Fig 4: The logarithm of the collapse time $\tau$ of a square vertical 
domain plunged into the horizontal SAF phase is plotted versus
the logarithm of the length $L$ of its side. 
Black squares are Monte Carlo data
obtained by averaging over $10$ different trials in the zero temperature case
with parameters $J_1=0.1$, $J_2=-1$ and $J_3=0$. The slope of the solid
line is $2$. Similar results have been obtained when different background
phases were considered. 
\vskip 0.5 cm
\par\noindent
Fig 5: Scaling collapse of the correlation functions ${\Gamma}_{\ell}(r,t)$ and 
${\Gamma}_t(r,t)$ achieved by plotting them versus $r/t^{0.5}$.
Results are given for $\beta=\infty$, $J_2=-1$, $J_3=0$, $J_1=0.1$ above and
$J_1=-0.1$ below; we averaged over $50$ different histories on a  
$400\times 400$ system. Black (empty) circles, squares and triangles
correspond respectively to the transverse (longitudinal) correlation
functions measured at times $300$, $400$ and $500$. Solid lines are the
best OJK fits. 
\vskip 0.5 cm
\par\noindent
Fig 6: Scaling collapse of the correlation functions $C_{\ell}(r,t)$ and
$C_t(r,t)$ achieved by plotting them versus the scaling variable
$r/t^{1/3}$. Results are given for $\beta=1$, $J_1=0.4$, $J_2=-1$ and 
$J_3=0$; we averaged Monte Carlo data over $50$ different stories on
a $800\times 800$ system. Both the longitudinal and transverse correlation
functions are shown at times $68000$ ($\circ$), $70000$ ($\square$) and
$72000$ ($\triangle$).

\newpage
\begin{figure}
\vskip -10mm
\inseps{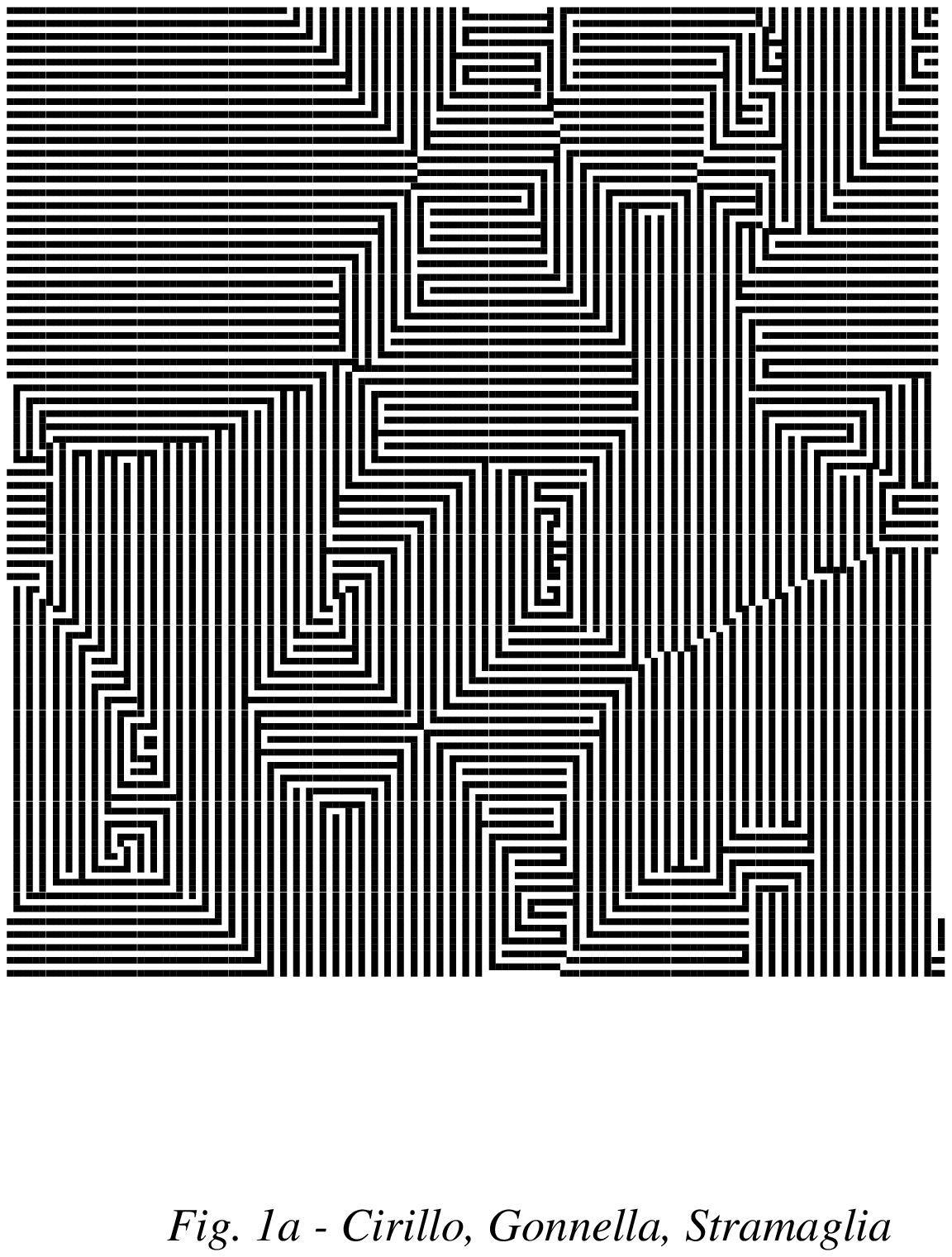}{1}
\vskip -10mm
\label{fig:1a}
\end{figure}

\newpage
\begin{figure}
\vskip -10mm
\inseps{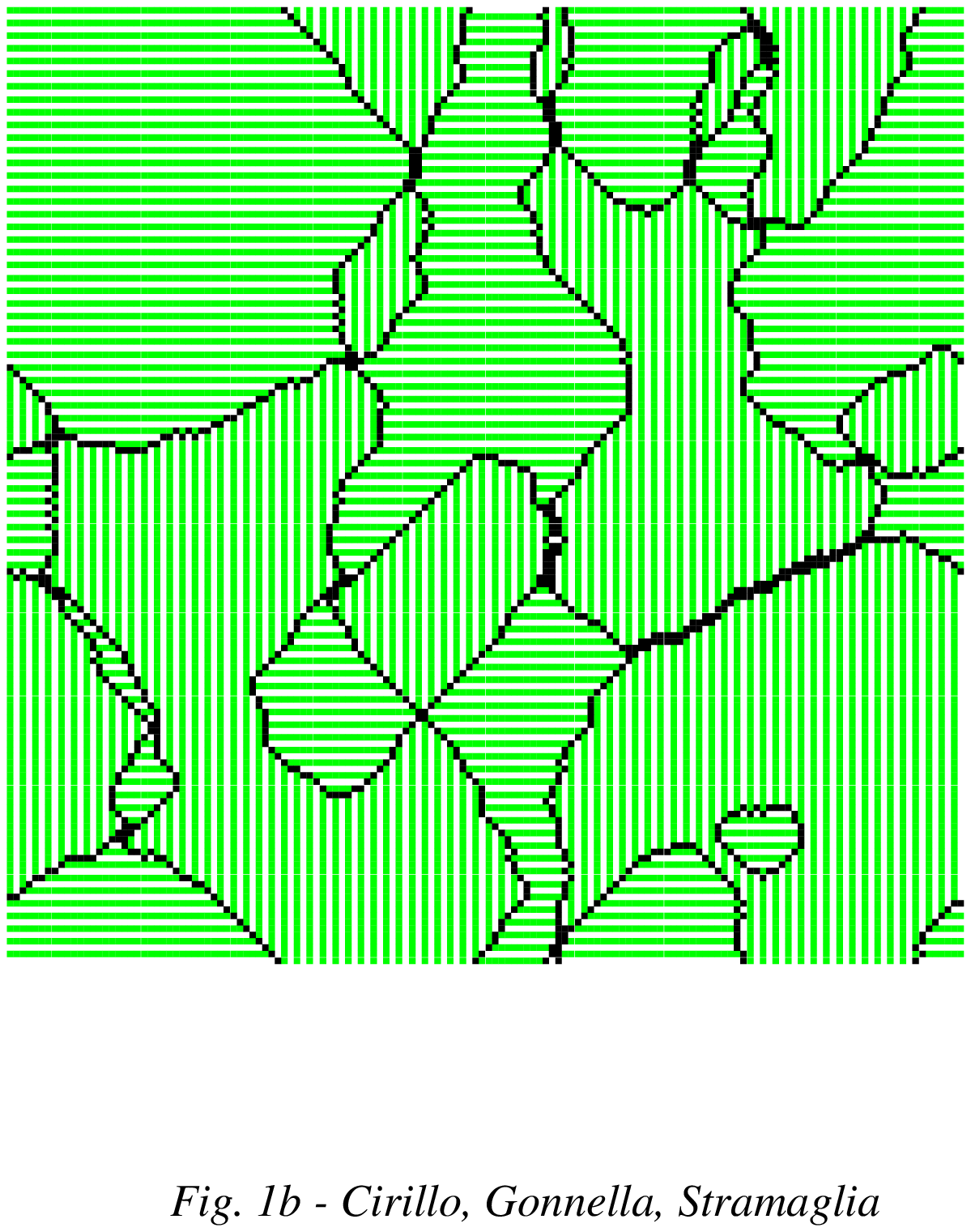}{1}
\vskip -10mm
\label{fig:1b}
\end{figure}

\newpage
\begin{figure}
\vskip -10mm
\inseps{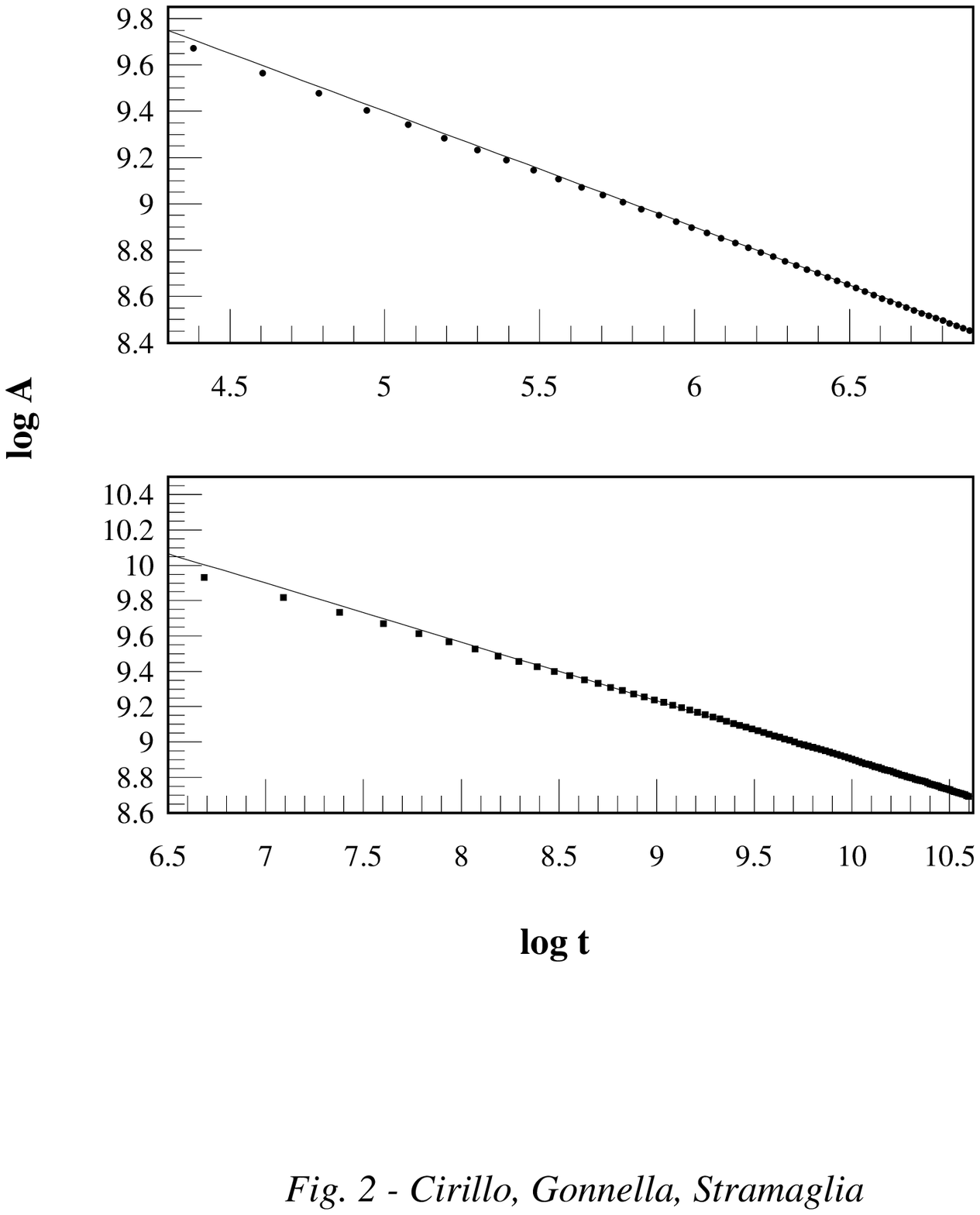}{1}
\vskip -10mm
\label{fig:inter}
\end{figure}

\newpage
\begin{figure}
\vskip -10mm
\inseps{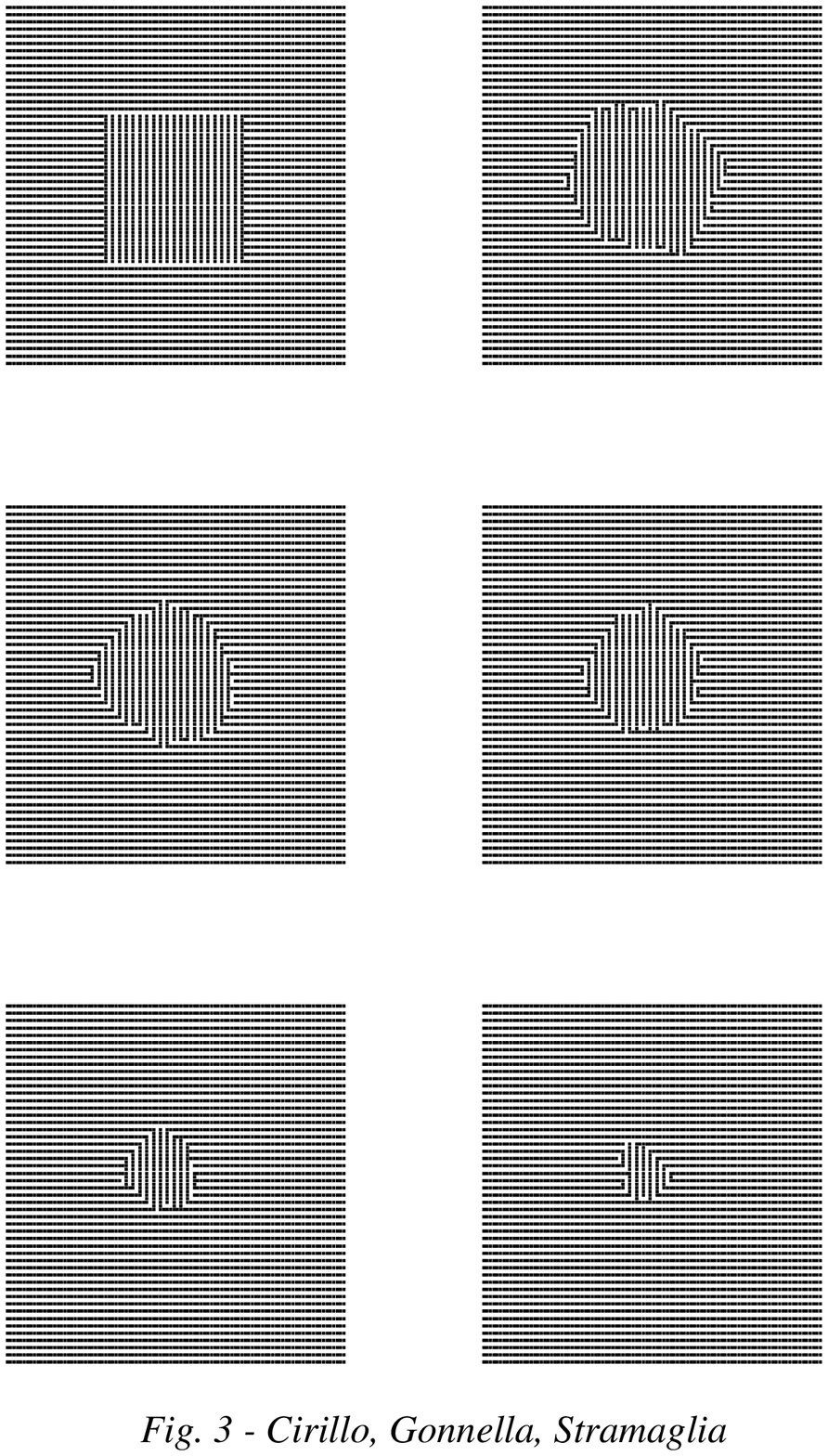}{1}
\vskip -10mm
\label{fig:shrink}
\end{figure}

\newpage
\begin{figure}
\vskip -10mm
\inseps{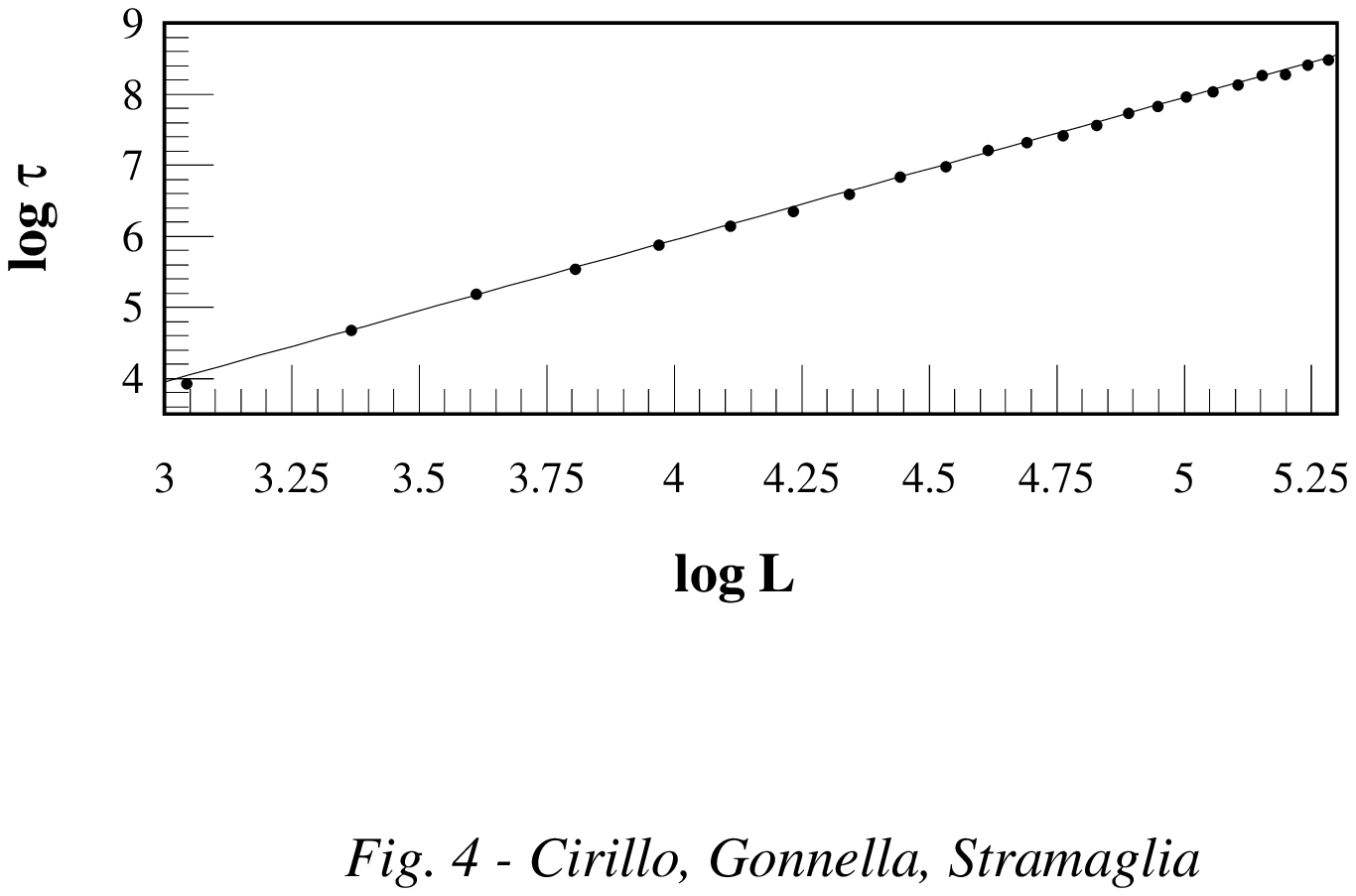}{1}
\vskip -10mm
\label{fig:ball}
\end{figure}

\newpage
\begin{figure}
\vskip -10mm
\inseps{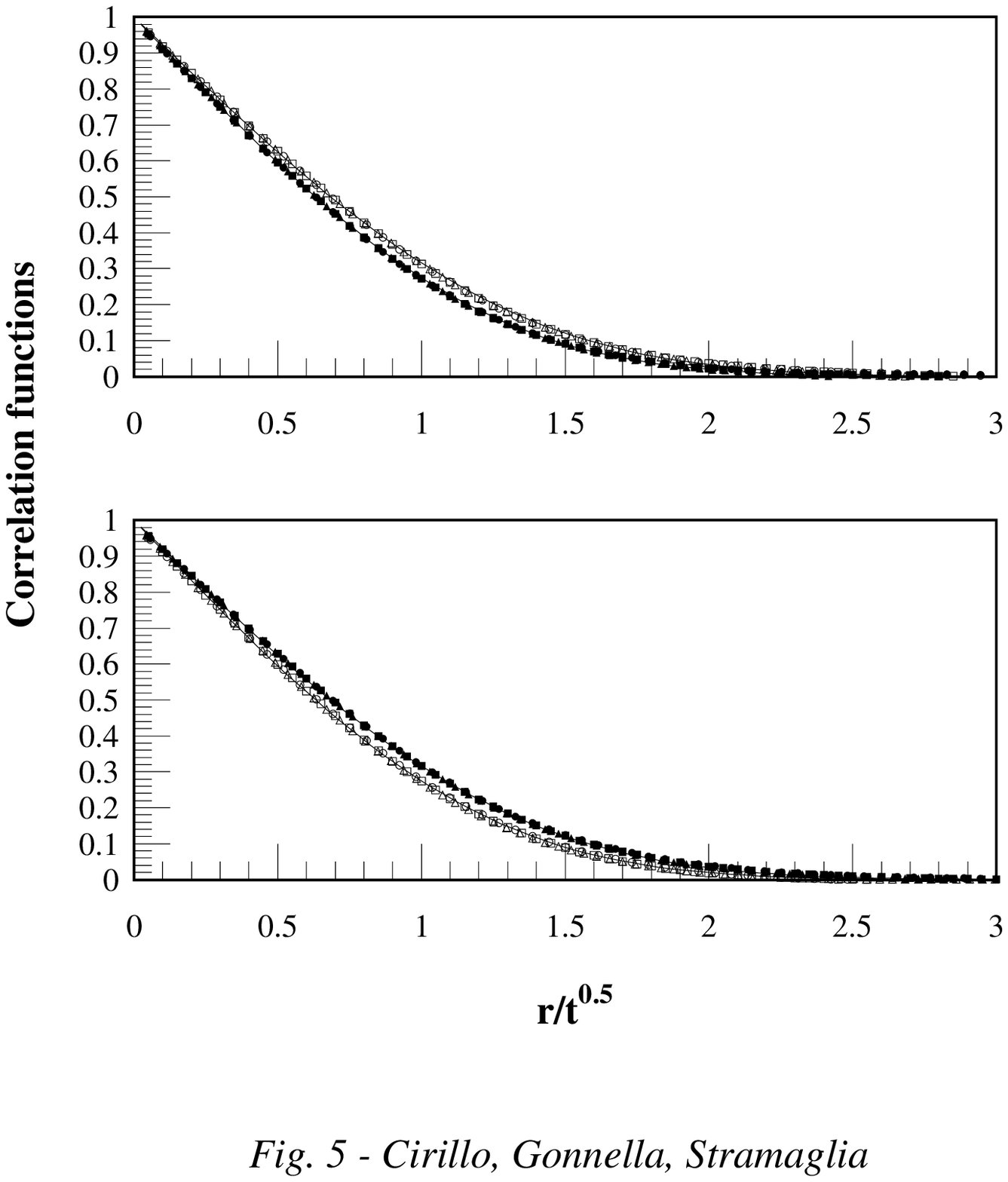}{1}
\vskip -10mm
\label{fig:bind}
\end{figure}

\newpage
\begin{figure}
\vskip -10mm
\inseps{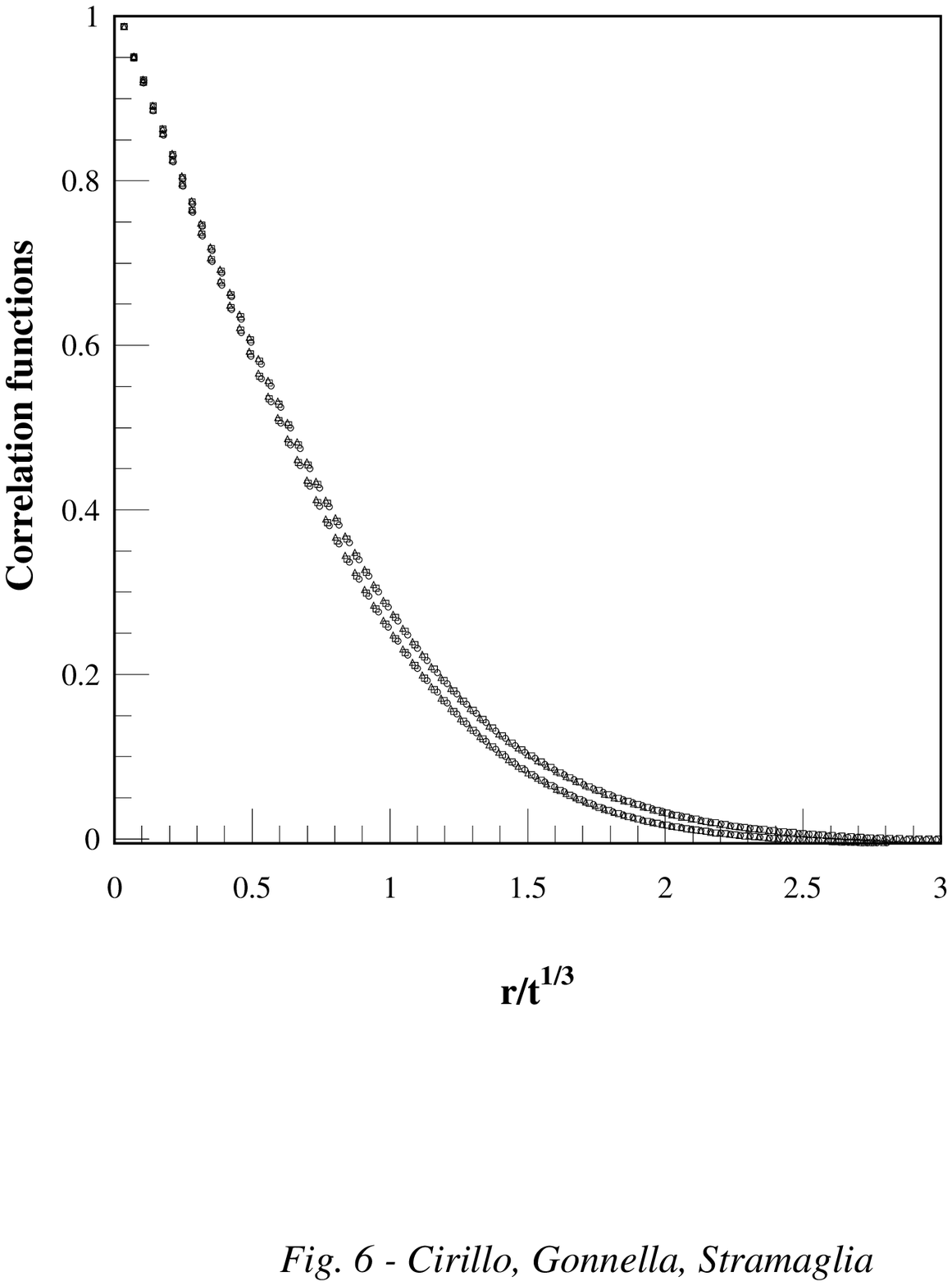}{1}
\vskip -10mm
\label{fig:kaw}
\end{figure}


\newpage
\begin{thebibliography}{99}
\bibitem{BR} A.J. Bray, Adv. Phys. {\bf 43}, 357 (1994). 
\bibitem{G} J.D. Gunton et al., in 
{\it Phase Transitions and Critical Phenomena}, edited by C. Domb and J.L. 
Lebowitz (Academic, New York, 1983), Vol.8.
\bibitem{B} S.F. Bates, Science {\bf 251}, 898 (1991).
\bibitem{LU} I.M. Lifshitz, Zh. Eskp. Teor. Fiz. {\bf 42}, 1354 (1962); S.M. 
Allen, J.W. Cahn, Acta. Metall. {\bf 27}, 1085 (1979).
\bibitem{LD} I.M. Lifshitz, V.V. Slyozov, J. Chem. Solids {\bf 19}, 
35 (1961).
\bibitem{Leb} M. Rao, M.H. Kalos, J.L. Lebowitz, J. Marro, Phys. Rev. B
{\bf 13}, 4328 (1976); J.F. Marko, J.T. Barkema, Phys. Rev. E {\bf 52}, 
2522 (1995).
\bibitem{SH} J.D. Shore, M. Holzer, J.P. Sethna, Phys. Rev. B {\bf 46}, 
11379 (1992); M. Rao, A. Chakrabarti, Phys. Rev. E {\bf 52}, R13 (1995). 
\bibitem{BAX} R.J. Baxter, {\it Exactly Solved Models in Statistical 
Mechanics} (Academic Press, London, 1982).
\bibitem{SB} A. Sadiq, K. Binder, Journ. Stat. Phys. {\bf 35}, 517 (1984).
\bibitem{CGS} E.N.M. Cirillo, G. Gonnella, S. Stramaglia, in press on 
Phys. Rev. E.
\bibitem{RU} A.D. Rutenberg, Phys. Rev. E {\bf 54}, R2181 (1996).
\bibitem{HB} K. Humayun, A. Bray, J. Phys. A: Math. Gen. {\bf 24}, 1915 (1991).
\bibitem{GLA} R.J. Glauber, Jour. Math. Phys. {\bf 4}, 294 (1962).
\bibitem{KAW} K. Kawasaki, in {\it Phase Transitions and Critical Phenomena}, 
Vol.2, edited by C. Domb and M. Green (Academic Press, London, 1970).
\bibitem{zero} The spin is
flipped with probability $1$ if $\Delta H < 0$, with probability $0.5$ if
$\Delta H$ vanishes and is not flipped if $\Delta H$ is positive; $\Delta H$ is
the variation of the hamiltonian corresponding to the spin flip.
\bibitem{OJK} T. Ohta, D. Jasnow, K. Kawasaki, Phys. Rev. Lett. {\bf 49},
1223 (1982).
\end{thebibliography}
\end{document}